\documentclass[journal]{IEEEtran}
\ifCLASSINFOpdf
\else
\fi
\usepackage[dvipdfmx]{graphicx}
\usepackage{amsmath,amssymb,amsthm}
\usepackage{algorithmic,algorithm}
\usepackage{bm}
\usepackage{comment}
\usepackage{array}
\usepackage{multirow}
\usepackage{cite}
\usepackage{subcaption}
\usepackage{tabularx}
\usepackage{enumerate}
\usepackage{empheq}
\usepackage{color}
\usepackage{booktabs}

\allowdisplaybreaks 

\theoremstyle{definition}

\newtheorem{problem}{\textbf{Problem}}
\newtheorem{note}{\textbf{Remark}}

\newcommand{\req}[1]{\eqref{#1}} 
\newcommand{\reqs}[1]{\eqref{#1}} 
 
\newcommand{\rfig}[1]{Fig.~\ref{#1}} 
\newcommand{\rfigs}[1]{Figs.~\ref{#1}}
\newcommand{\rtab}[1]{Table~\ref{#1}}

\renewcommand{\figurename}{Fig.}
\renewcommand{\tablename}{Table}
\newcommand{\mini}{\mathop{\rm minimize}\limits}

\begin{document}
\title{Trajectory Planning of Weakly Supervised Aircraft}

\author{Sho~Yoshimura~and 
        Masaki~Inoue,~\IEEEmembership{Member,~IEEE}
\thanks{S. Yoshimura and M. Inoue are with the Department of Applied Physics and Physico-Informatics, Keio University, Yokohama, Japan (e-mail: minoue@appi.keio.jp).}}

\markboth{Journal of \LaTeX\ Class Files,~Vol.~14, No.~8, August~2015}%
{Shell \MakeLowercase{\textit{et al.}}: Bare Demo of IEEEtran.cls for IEEE Journals}

\maketitle

\begin{abstract}
In this paper, we propose a novel framework of air traffic management (ATM).
The framework is in particular characterized by the trajectory planning of \emph{weakly supervised} aircraft; the air traffic control (ATC) does not completely determine the trajectory of each aircraft unlike conventional planning methods, but determines \emph{allowable safe sets} of trajectories. 
ATC requests pilots to select their own trajectories from the sets, and the pilots determine ones by pursuing their own aims.
For example, the selection can be based on pilot preferences and airline strategies.
This \emph{two stage} ATM system contributes to simultaneously achieve the both objectives of the ATC and pilots such as fuel cost minimization under safety management.
The effectiveness of the proposed ATM system is demonstrated in a simulation using actual air traffic data.
\end{abstract}

\begin{IEEEkeywords}
Trajectory planning, air traffic management, optimization
\end{IEEEkeywords}

\IEEEpeerreviewmaketitle

\section{Introduction}
\IEEEPARstart{W}{ith} the growth of air traffic all over the world, air traffic management (ATM) systems need to be further developed and to be operated more efficiently.
As reported in \cite{ICAO}, ATM is defined as dynamic and integrated management of air traffic and airspace through the provision of facilities and seamless services in collaboration with parties.
In this paper, we focus only on control and management problems of aircraft trajectories from departure to arrival.

\begin{figure}[b]
\centering
\includegraphics[scale=0.6]{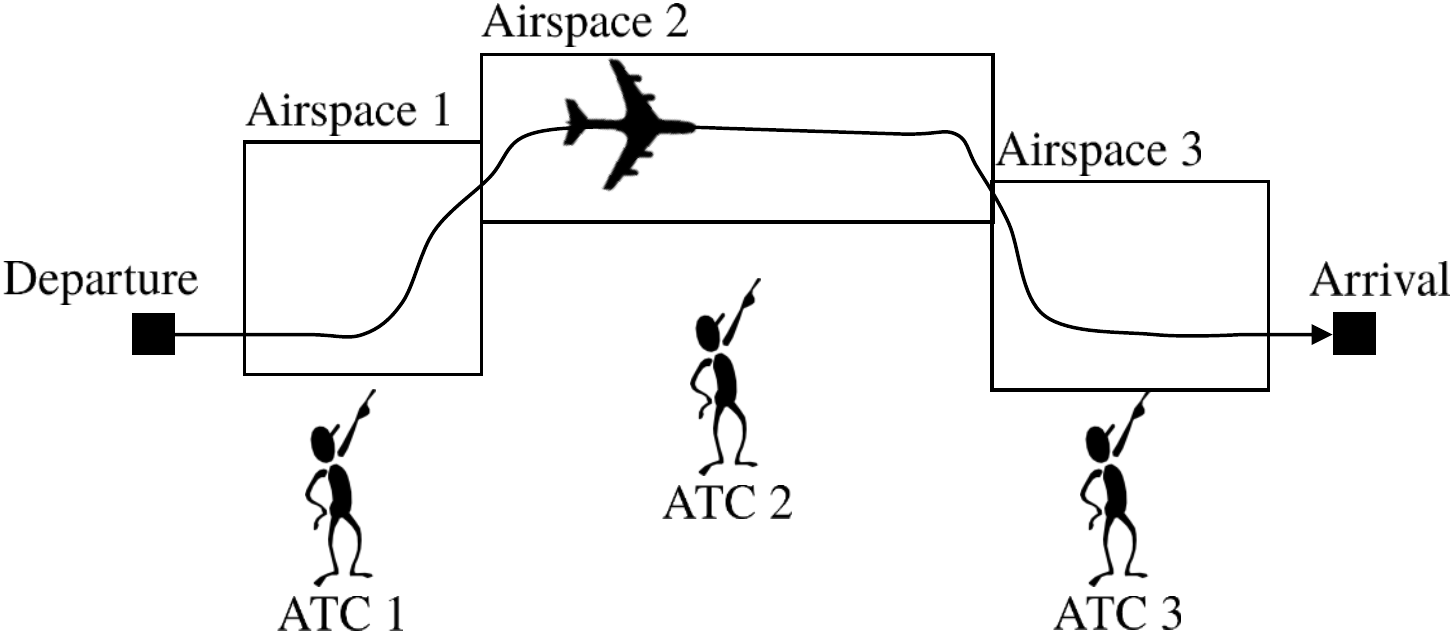}
\caption{Illustration of the current ATM. Control tasks of aircraft from departure to arrival are divided between airspace. For each airspace, an air traffic controller designs aircraft trajectories manually and provides them to pilots.\label{ATCcurrent}}
\end{figure}

We review the current ATM system and find its drawback.
A sketch of the current system is given in \rfig{ATCcurrent}.
As illustrated in the figure, overall airspace is divided to some smaller spaces.
The current ATM is operated in a decentralized manner; in other words, a control task corresponding to each divided space is assigned to each air traffic control (ATC).
Then, each ATC is responsible for safety of aircraft in an assigned space and manually designs aircraft trajectories.
The design is based on positions of aircraft that are obtained from radar information.  
The trajectories designed by ATC are provided to pilots.
A main limitation of the current ATM system lies in manually designing trajectories by ATC.
With the growth of air traffic, the number of aircraft contained in airspace increases.
This increases burdens for the corresponding ATC, which inhibits efficient trajectory design.
In addition, lack of communication between ATCs, caused by the traffic growth, can further deteriorate the efficiency. 
Therefore, a computer-aided and human-assist ATM system where overall airspace is managed in a centralized manner, is required.

In the past two decades, various computer-aided methods of generating aircraft trajectories have been proposed\cite{miyazawa2013dynamic,pallottino2002conflict,bonami2013multiphase,
menon1999optimal,han2017traffic,tang2016optimal,
vela2010near,hu2002optimal,bicchi2000optimal,toratani2018effects}.
One of the main topics is conflict avoidance.
Please see, e.g., \cite{menon1999optimal,han2017traffic,tang2016optimal,
vela2010near,hu2002optimal,bicchi2000optimal}.
The conflict avoidance is involved in optimal trajectory generation problems\cite{tang2016optimal,vela2010near,hu2002optimal,
bicchi2000optimal,menon1999optimal}.
In the current ATM system and previous works\cite{pallottino2002conflict,menon1999optimal,han2017traffic,
vela2010near,hu2002optimal,bicchi2000optimal,tang2016optimal,
toratani2018effects}, ATC determines aircraft trajectories completely and provides them to pilots.
Pilots must obey the provided trajectories.

The priority to ATC in trajectory determination plays a role for safety management of aircraft.
In addition to the safety aim, ATC can achieve to improve airport usage, e.g., by maximizing airport throughput.
We can say that the current ATM system meets aims of ATC.
On the other hand, the priority to ATC may not be positively acceptable for pilots and airlines.
Trajectories determined and enforced by ATC may be fuel-consuming or delayed ones, which are contrary to the aims of pilots and airlines.
This paper addresses a novel ATM system design where trajectories are determined based on pilots and airlines aims in addition to safety constraints.

\begin{figure}
\centering
\includegraphics[scale=0.6]{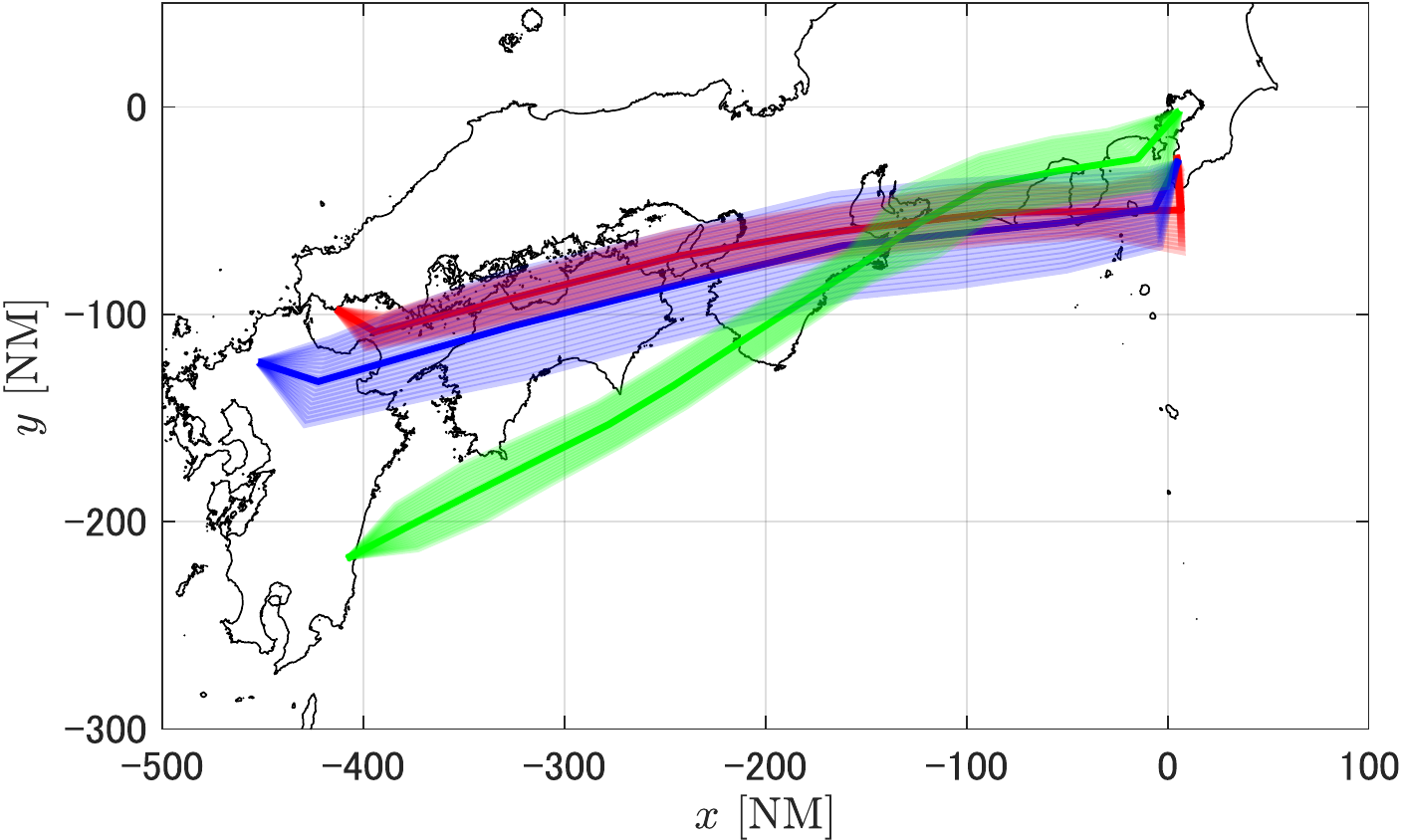}
\caption{Example of all trajectories that satisfy ATC requirements including conflict avoidance.
The three solid lines represent actual trajectories.
The translucent solid lines represent trajectories that avoid aircraft conflicts. Here, the conflict avoidance is defined that aircraft-aircraft distance is longer than 3 NM.
\label{motivation}}
\end{figure}

To show the feasibility of the ATM system design, we illustrate the existence of degree of freedom (DOF) in actual aircraft trajectories determined by ATC.
Actual trajectories are depicted in \rfig{motivation}, where trajectory data is extracted from CARATS (Collaborative Actions for Renovation of Air Traffic Systems) Open Data\footnote{CARATS Open Data includes 3D-position and time data of all IFR (Instrument Flight Rules) commercial flights in Japanese airspace.} \cite{MLIT2010,matsuda2017arrival,fukuda2015air,wickramasinghe2017feasibility} on May 11, 2015.
In the figure, the three solid lines represent actual trajectories.
The translucent lines represent all trajectories where conflict avoidance is achieved.
The DOF in trajectories, illustrated by the translucent lines, implies that the aims of pilots and airlines can be pursued in addition to safety constraints.
Motivated from this fact, this paper studies a novel framework of ATM systems by utilizing this DOF in trajectory design.

\begin{figure}
\centering
\includegraphics[scale=0.55]{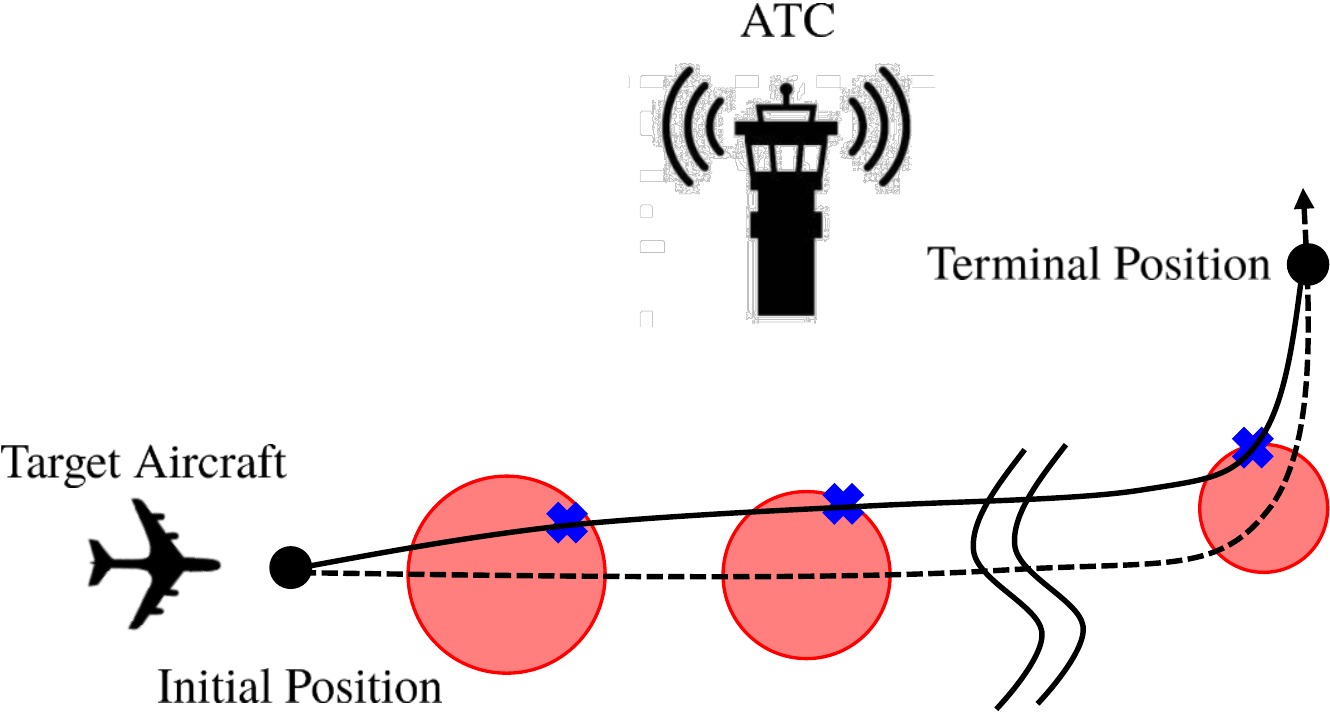}
\caption{Illustration of the proposed ATM framework.
In this figure, the black points represent the initial and terminal positions of the target aircraft.
ATC designs an \emph{allowable safe set} of trajectories which is represented by the red disks.
A trajectory selected from the set by the pilot is represented by the blue cross marks.
\label{framework}}
\end{figure}

In this paper, we propose a framework of ATM systems where ATC \emph{weakly supervises} aircraft for the trajectories planning.
An illustration of the proposed framework is shown in \rfig{framework}.
In the ATM system, ATC designs \emph{allowable safe sets} of aircraft trajectories, where conflict avoidance and other aims of ATC are achieved, as depicted by red disks in the figure.
Then, ATC requests pilots to select their trajectories from the sets.
The selection can be based on pilot preferences or airline strategies.
It is noted that for the trajectory selection, communication between pilots, such as cooperation, competition and negotiation, is not required.
Each pilot individually determines his/her trajectory by pursuing his/her own aims, 
e.g., by minimizing fuel costs or by reducing uncertainty of arrival time in a decentralized manner.

Designing \emph{trajectory-sets} for aircraft has been also studied as 
flow corridors, corridors-in-the-sky, or airspace tubes 
\cite{hoffman2008principles,takeichi2016benefit,xue2009design,yousefi2013dynamic}.
The concept of the corridor means an exclusive lane designed in airspace where autonomy is allowed for each aircraft; each pilot can design his/her trajectory within the lane without instructions by ATC.
This implies that the pilot must be responsible for avoiding an air miss or conflict with other aircraft by communicating with other pilots.
This drawback is overcome in the proposed ATM system; 
each pilot needs not to address the conflict avoidance problem, but 
simply pursues his/her aims based on a given trajectory-set. 

The rest of this paper is organized as follows.
In Section I\hspace{-.1em}I, the proposed framework of ATM systems and problems of trajectory planning are briefly stated.
In Section I\hspace{-.1em}I\hspace{-.1em}I, a trajectory design problem addressed by ATC is formulated as an optimization problem.
By solving the problem, trajectory-sets that avoid aircraft conflicts are obtained.
In Section I\hspace{-.1em}V, in a similar manner to Section I\hspace{-.1em}I\hspace{-.1em}I, a trajectory design problem addressed by pilots is formulated as an optimization problem.
At the end of the section, the proposed ATM system is reviewed.
In Section V, we show the effectiveness of the proposed system in a numerical simulation with CARATS Open Data. 

\section{Weak Control in ATM}
\subsection{Outline of Proposed ATM}
\begin{figure}
\centering
\includegraphics[scale=0.5]{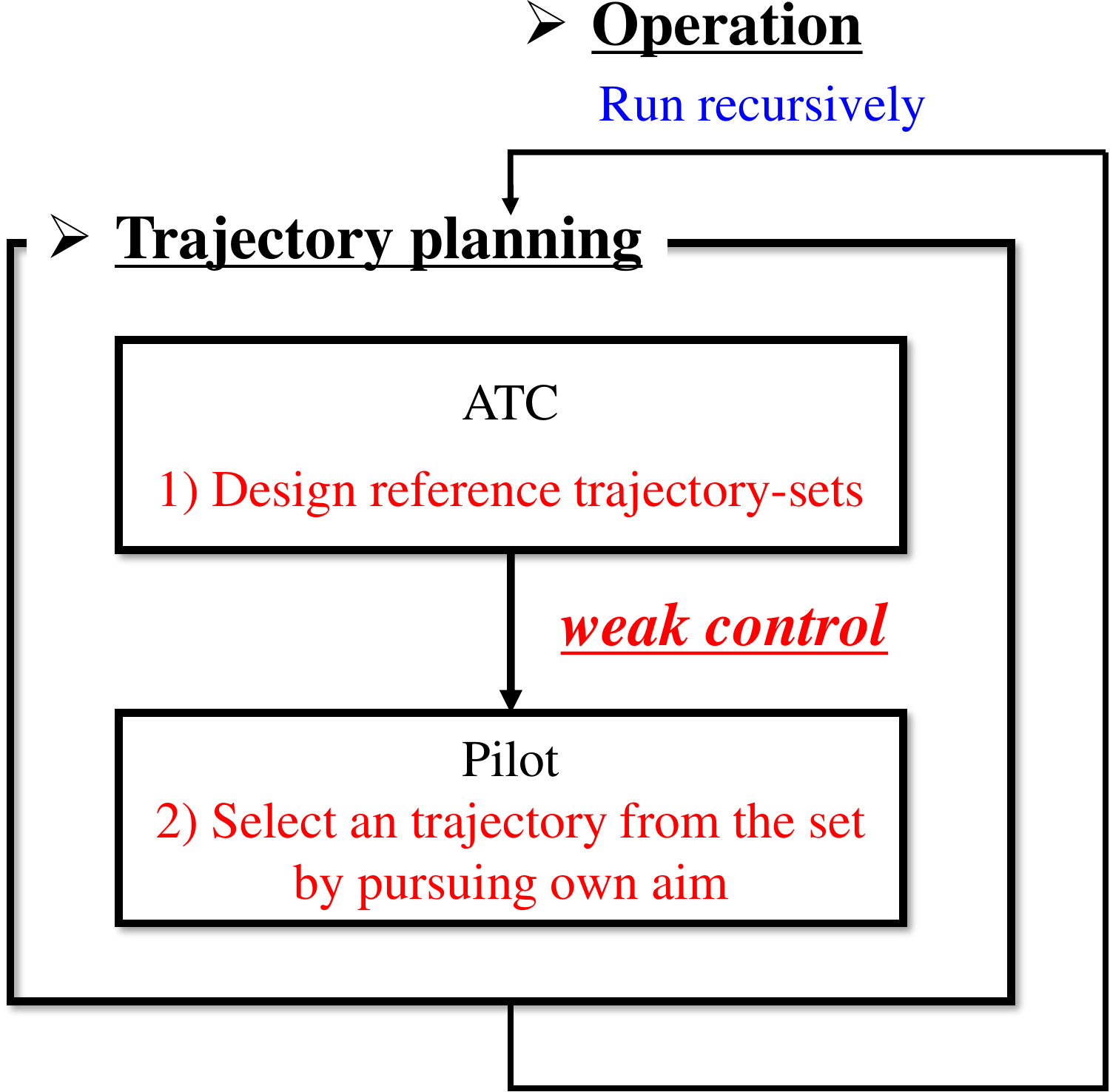}
\caption{Operation flow of the proposed ATM system.
Trajectory planning runs recursively at some time interval.
ATC designs \emph{allowable safe sets} of trajectories, and then requests them to the pilots.
Pilots select trajectories from the sets independently of each other.
The selections are based on their own aims.
\label{concept}}
\end{figure}

In this subsection, we briefly show the outline of the proposed ATM system.
The operation flow of the ATM system is given in \rfig{concept}.
Operation of the ATM system is to control aircraft from departure to arrival.
Trajectory planning runs recursively at some time interval.
For each trajectory planning, ATC designs \emph{allowable safe sets} of trajectories, and then requests pilots to select their trajectories from the sets.
The selection by pilots is performed independently of each other.
This two-stage trajectory planning achieves aims of pilots and airlines such as minimization of fuel consumption, while satisfying requirements by ATC such as conflict avoidance. 
 
Recall that the requests from ATC to pilots are given by \emph{allowable safe sets} of trajectories.
The sets are utilized for pilots to minimize fuel costs based on detailed aircraft models and weather conditions.
We can say that in the proposed ATM system pilots are \emph{weakly supervised} by ATC.
The weakness contributes to \emph{separate} ATC management and pilots optimization;  
in other words, safe management is performed independently of pilots choices.

\subsection{Problem Formulation}
Notation utilized in this paper is listed on \rtab{tab:variables}.
In this subsection, problems of trajectory planning are briefly stated.
Let $N$ be the number of target aircraft considered in the problem.
In this paper, trajectories refer to positions of aircraft at equal time intervals.
Trajectories are restricted to two dimensions for simplicity, but they are easily extended to three dimensions.
The position of aircraft $i\in \mathcal{N}:= \{1,\ldots,N\}$ at time $k\in\{0,1,\ldots \}$ is denoted by $C_i(k):=[\,x_i(k)~y_i(k)\,]^{\top}$.

Variables of aircraft are defined below.
Let $v_i(k)$, $\theta_i(k)$, $u_i(k)$, and $\psi_i(k)$ be the speed, angle, speed difference, and angle difference of aircraft $i$ at time $k$, respectively.
The state of aircraft $i$ at time $k$ is denoted by $X_i(k):=[\,C_i^{\top}(k)~v_i(k)~\theta_i(k)\,]^{\top}$.

As stated above, each trajectory planning is composed of two stages 1) trajectories design and request by ATC side and 2) trajectory design by pilots side.
First, we focus on the design problem addressed by ATC as follows.
We assume that the initial time $t_i$, the terminal time $T_i$, the initial state $X_i(t_i)$, and the terminal state $X_i(T_i)$ for all aircraft $i\in \mathcal{N}$ are given.
Then, ATC designs reference trajectory-sets $\mathcal{W}_i(k),~k\in\{t_i+1,\ldots, T_i-1\},~i\in \mathcal{N}$.
Let $\mathcal{W}_i(k)$ be a disk region.
Then, the problem is reduced to a problem of finding the center $C_{i}(k)$ and radius $r_i(k)$ of $\mathcal{W}_i(k)$. 
In the remainder of this paper, $\mathcal{W}_i(k)$ is described as $\mathcal{W}_i(k)=\{ C_{i}(k),r_i(k)\}$.
Through the trajectory design, ATC pursues the following aims;
\begin{itemize}
\item maximizing the DOF in the trajectory-set, defined by $\| r_i\|$,
\item guaranteeing safety of aircraft by conflict avoidance.
\end{itemize}
For simplification of notation, the set of time of aircraft $i$ is denoted by $\mathcal{T}_i(t_i):=\{t_i+1,\ldots,T_i-1\}$.

Letting $r:=[\,r_1\, \cdots \, r_N\,]^{\top}$, the trajectory design problem addressed by ATC is summarized as follows
\begin{problem}\label{prob_atc}
For given $X_i(t_i)$, $X_i(T_i)$, $i\in \mathcal{N}$, find $\mathcal{W}_i(k)$, $k\in \mathcal{T}_i(t_i)$, $i\in \mathcal{N}$ maximizing $\|r\|$ subject to some constraints. 
\end{problem}
\noindent
\textbf{Problem~1} corresponds to the design problem 1), which is illustrated in the operation flow of \rfig{concept}.

Next, we briefly mention a trajectory design problem addressed by pilots, which is solved next to \textbf{Problem 1}.
Each pilot solves the problem independently of each other.
Assume that the initial state $X_i(t_i)$, the terminal state $
X_i(T_i)$, the requested trajectory-set $\mathcal{W}_i(k)$, $k\in \mathcal{T}_i(t_i)$, and disturbances $d_i(k)$, $k\in \mathcal{T}_i(t_i)$ are given.
Then, each pilot selects a trajectory $\hat{C}_i(k)$, $k\in \mathcal{T}_i(t_i)$ by pursuing his/her aims, e.g., minimizing a fuel cost.
The trajectory design problem addressed by a pilot labeled by $i$, is summarized as follows.
\begin{problem}\label{prob_pilot}
For given $X_i(t_i)$, $X_i(T_i)$, $\mathcal{W}_i(k)$, $k\in \mathcal{T}_i(t_i)$, find $\hat{C}_i(k)$, $k\in \mathcal{T}_i(t_i)$ minimizing some costs.
\end{problem}
\noindent
\textbf{Problem~2} corresponds to the design problem 2), which is illustrated in the operation flow of \rfig{concept}. 

\begin{note}
It is assumed that model information of aircraft and a weather condition are available for each pilot, while they are not available for ATC.
Precise model information and a weather condition play important role of reducing a fuel cost.
Therefore, the pilot utilizes such information to solve \textbf{Problem 2} in an efficient manner and to reduce the achievable cost.
\end{note}

\begin{table}
\centering
\caption{Definition of variables of aircraft $i$ in this paper.\label{tab:variables}}
\begin{tabular}{ll}\toprule
Variables & Description \\ \hline \hline
$N$ & number of target aircraft\\ \hline
$t_i$ & initial time\\ \hline
$T_i$ & terminal time\\ \hline
$\mathcal{W}_i$ & reference trajectory-set (region defined as disk)\\ \hline
$C_{i}$ & center of $\mathcal{W}_i$, i.e., $C_i = [\,x_i,\, y_i\,]^{\top}$ \\ \hline
$r_i$ & radius of $\mathcal{W}_i$\\ \hline
$C_{\mathrm{sta},i}$ & standard trajectory \\ \hline
$\Delta$ & deviation between standard trajectory and center of region\\ \hline
$\hat{C}_i$ & trajectory to be designed by pilot \\ \hline
$C_{\mathrm{pre},i}$ & trajectory pilot select at the last trajectory planning\\ \hline
$d_i$ & disturbance\\ \hline
$x_i$ & x coordinate of position \\ \hline
$y_i$ & y coordinate of position \\ \hline
$v_i$ & speed \\ \hline
$\theta_i$ & angle\\ \hline
$u_i$ & speed difference \\ \hline
$\psi_i$ & angle difference \\ \hline
$X_i$ & state, i.e., $X_i = [\,C_i^{\top},v_i,\theta_i\,]^{\top}$ \\ \bottomrule
\end{tabular}
\end{table}

\section{Trajectory Design by ATC}
In this section, we mathematically formulate the trajectory design problem addressed by ATC.
\textbf{Problem~\ref{prob_atc}}, which is briefly stated in Section I\hspace{-.1em}I, is re-formulated as an optimization problem.

\subsection{Cost Function}
In this subsection, cost functions of the trajectory design problem for ATC are defined.
The decision variables are speed differences $u:=[\,u_1\cdots u_N\,]^{\top}$, angle differences $\psi:=[\,\psi_1\cdots \psi_N\,]^{\top}$, and radii $r$ of all aircraft.
Trajectory-sets to be designed are evaluated by the size of $r$ and the deviations from a standard trajectory $C_{\mathrm{sta},i}(k)$, $k\in \mathcal{T}_i(t_i)$.
The cost function to evaluate $r$ is given by
\begin{align}
\label{eq:J1}
J_1
=
-\sum_{i=1}^{N}\sum_{k=t_i+1}^{T_i-1}
\ln{\left(r_i(k)+\varepsilon\right)},
\end{align}
where $\varepsilon$ is a positive constant.
This role of $\varepsilon$ is to prevent divergence of $J_1$.
In addition, the cost function to evaluate the deviations is given by
\begin{align}
J_2
=
\sum_{i=1}^{N}\sum_{k=t_i+1}^{T_i-1}
\Delta_i(k)^2
+
\sum_{i=1}^{N}\sum_{k=t_i+1}^{T_i-2}
\left(
\Delta_i(k+1)-\Delta_i(k)
\right)^2,
\end{align}
where $\Delta_i(k)=C_i(k)-C_{\mathrm{sta},i}(k)$.
The role of the second term is to prevent a large angle change in one time step.
By this $J_2$, we aim at generating realistic trajectories, while avoiding impractical ones.
Then, the cost function utilized in the problem here is defined  by 
\begin{align}
\label{eq:atc_cost}
J_{\mathrm{ATC}}(u,\psi,r) := 
J_1 + \alpha J_2,
\end{align}
where $\alpha$ is a non-negative constant.
\begin{note}
There is another candidate of the cost function $J_1$ for evaluating the radii $r_i$.
For example, $\ln{\left(r_i(k)+\varepsilon\right)}$ in \req{eq:J1} can be replaced by $1/(r_i(k)+\varepsilon )^2$.
In this paper, numerically tractable $J_1$, given by \req{eq:J1}, is considered.  
\end{note}

\subsection{Constraints}
\begin{figure}[t]
\centering
\includegraphics[scale=0.5]{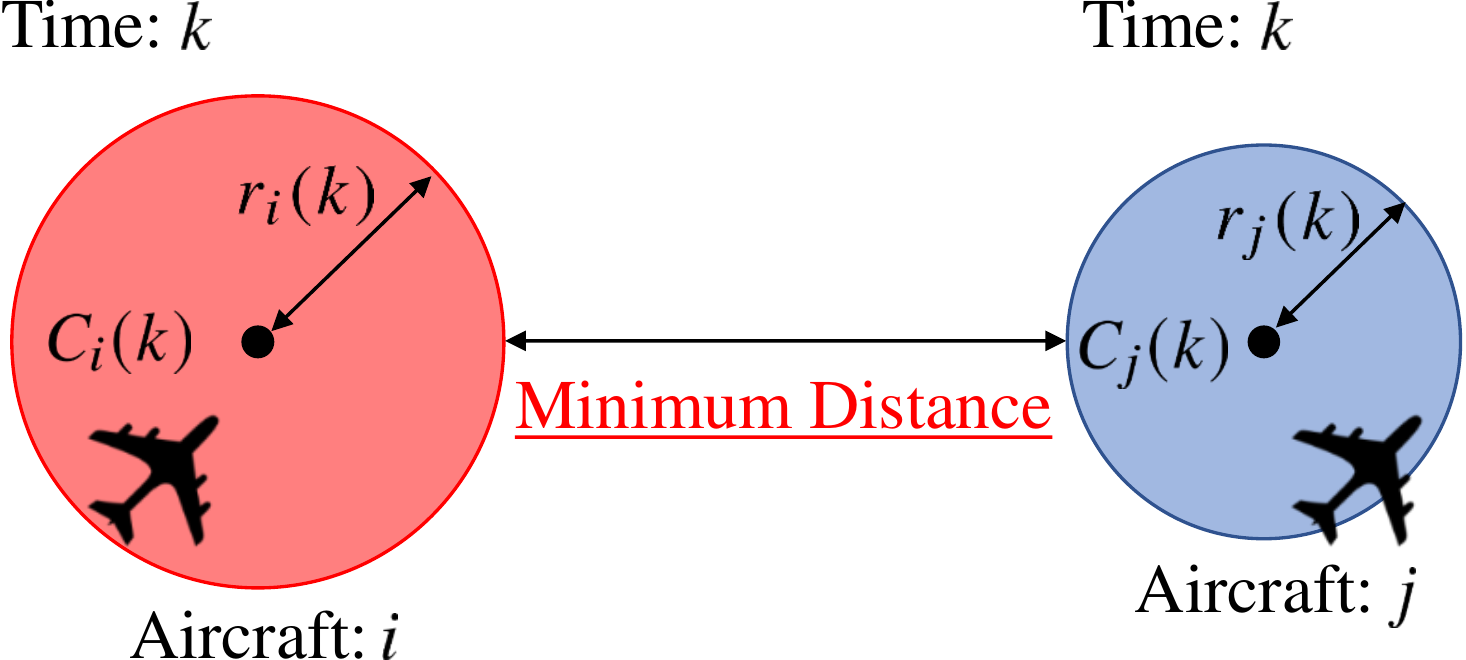}
\caption{Constraint for conflict avoidance. The minimum distance between the aircraft $i$ and $j$ at time $k$, which is described by the left side of \req{eq:conflict}, is shown.\label{fig:margin}}
\end{figure}

\begin{figure}
\centering
\includegraphics[scale=0.5]{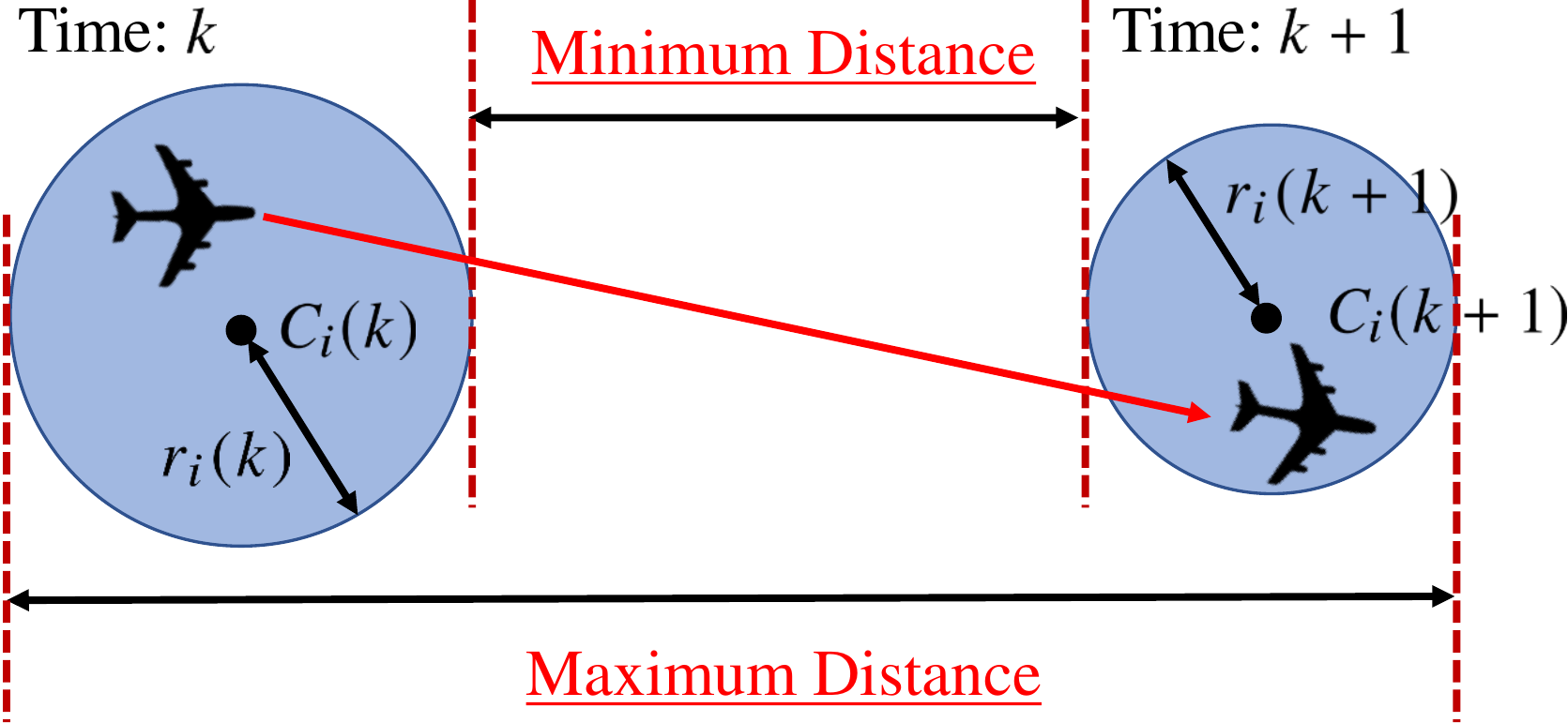}
\caption{Constraints for feasibility. The minimum and maxmum distance of adjacent regions at time $k$ and $k+1$, which are described by the left side of \req{eq:const_r_1} and \req{eq:const_r_2}, respectively, are shown.\label{fig:r_const}}
\end{figure}

\begin{figure}[t]
\centering
\includegraphics[scale=0.5]{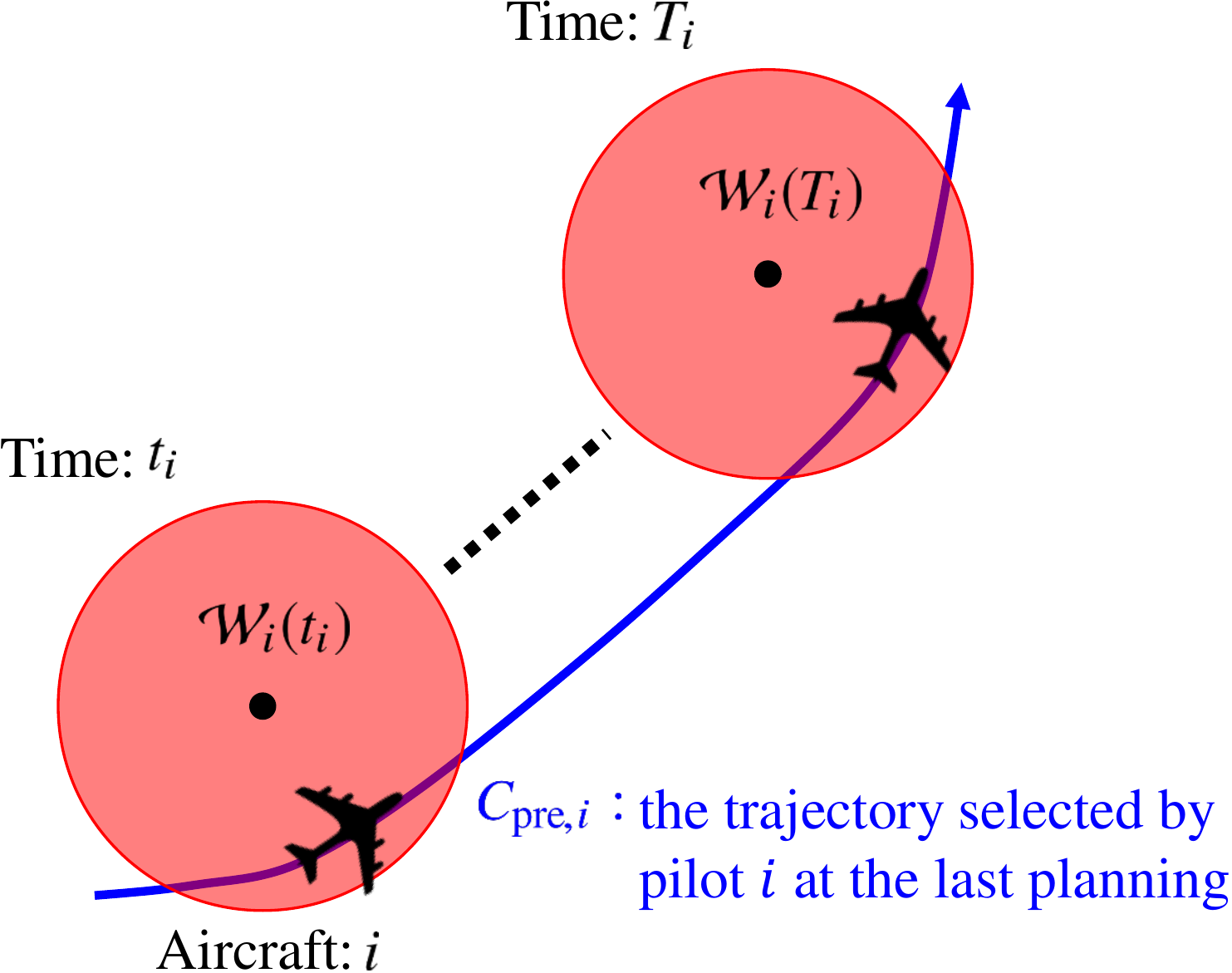}
\caption{Operation constraint.
The trajectory $C_{\mathrm{pre},i}$, which is selected at the last planning, must be included in $\mathcal{W}_i$, which is redesigned by ATC at the current planning.
The trajectory $C_{\mathrm{pre},i}$ is represented by the blue solid line, while the regions $\mathcal{W}_i$ is represented by the red disks.
\label{fig:operation}}
\end{figure}

We give constraints derived from physical properties of aircraft, from ATM system requirements, and for ATM system operation.
First, the constraints derived from the physical properties are given as follows.
\begin{itemize}
\item Aircraft dynamics\\
Assume that the dynamics of all the target aircraft are the same.
The dynamics of each aircraft is described by
\begin{align}
\label{eq:dynamics}
\begin{bmatrix}
x_i(k+1)\\
y_i(k+1)\\
v_i(k+1)\\
\theta _i(k+1)\\
\end{bmatrix}
=
f\left(
\begin{bmatrix}
x_i(k)\\
y_i(k)\\
v_i(k)\\
\theta _i(k)\\
\end{bmatrix}
\right)
+
\begin{bmatrix}
0\\
0\\
u_i(k)\\
\psi _i(k)\\
\end{bmatrix},
\end{align}
where $f$ is given by
\begin{align}
\label{eq:f}
f\left(
\begin{bmatrix}
x_i(k)\\
y_i(k)\\
v_i(k)\\
\theta _i(k)\\
\end{bmatrix}
\right)
:=
\begin{bmatrix}
x_i(k) + v_i(k)\cos \theta _i(k)\\
y_i(k) + v_i(k)\sin \theta _i(k)\\
v_i(k)\\
\theta_i(k)\\
\end{bmatrix}.
\end{align}
Here, recall that $u_i(k)$ and $\psi _i(k)$ are speed difference and angle difference at time $k\in \mathcal{T}_i(t_i)$ of aircraft $i\in \mathcal{N}$.
In the model \req{eq:dynamics}, they are the control input to aircraft. 

\item Angular difference\\
Angular difference of each aircraft trajectory, denoted by $\psi_i(k)$, is restricted as follows.
\begin{align}
\label{const_angle}
-\Psi \le \psi_i(k) \le \Psi, \quad \forall k\in\mathcal{T}_i(t_i),~\forall i\in\mathcal{N},
\end{align}
where $\Psi$ is the maximum angle difference of a trajectory.
\item Speed difference\\
Speed difference of each aircraft trajectory, denoted by $u_i(k)$, is restricted as follows.
\begin{align}
-U \le u_i(k) \le U, \quad \forall k\in\mathcal{T}_i(t_i),~\forall i\in\mathcal{N},
\end{align}
where $U$ is the maximum speed difference to be applied.
\item Speed\\
Speed of each aircraft, denoted by $v_i(k)$, is restricted as follows.
\begin{align}
\label{eq:velocity}
V_{\mathrm{min}}\le v_i(k) \le V_{\mathrm{max}},
\quad \forall k\in\mathcal{T}_i(t_i),~\forall i\in\mathcal{N},
\end{align}
where $V_{\mathrm{min}}$ and $V_{\mathrm{max}}$ are the lower and upper bounds of $v_i(k)$.
\end{itemize}
Next, the constraints related to ATM system requirements are given as follows.
\begin{itemize}
\item Terminal constraint\\
Constraints on the speed and angle at the terminal time $T_i$ are given.
The deviation between the speed at $T_i$ and the reference $V_{\mathrm{ter}}$ is less than or equal to $\delta_v$, and the deviation between angle at $T_i$ and the reference $\Theta_{\mathrm{ter}}$ is less than or equal to $\delta_{\theta}$.
They are described by
\begin{align}
\label{eq:term_velocity}
&-\delta_v \le v_i(T_i)-V_{\mathrm{ter}} \le \delta_v,\quad \forall i\in\mathcal{N},\\
\label{eq:term_angle}
&-\delta_{\theta} \le \theta_i(T_i)-\Theta_{\mathrm{ter}} \le \delta_{\theta},\quad \forall i\in\mathcal{N}.
\end{align}
\item Constraint for conflict avoidance\\
A constraint for conflict avoidance is given by
\begin{align}
\label{eq:conflict}
&\|
C_i(k)-C_j(k)
\|
-
\left(
r_i(k) + r_j(k)
\right)
\ge
D,\\
\nonumber
&\forall k\in \{ \mathcal{T}_i(t_i)\cap \mathcal{T}_j(t_j) \}
,~\forall i\neq j\in \mathcal{N},
\end{align}
where $D$ is a positive constant.
We illustrate the meaning of \req{eq:conflict} with \rfig{fig:margin}.
The left side of \req{eq:conflict} represents the allowable minimum distance between two aircraft, which is illustrated in \rfig{fig:margin}.
Therefore, $D$ plays a role of a safety margin.
\item Constraint for feasibility\\
In this paper, feasibility means that the trajectory design problem performed by pilots are feasible.
In other words, every pilot can select a trajectory from a trajectory-set requested by ATC.
This is described as
\begin{align}
\nonumber
&\| C_i(k+1)-C_i(k)\| -
\left(r_i(k+1)+r_i(k)\right) \\
\label{eq:const_r_1}
&\qquad \qquad \ge V_{\mathrm{min}},~\forall k\in\{t_i \cup \mathcal{T}_i\},~\forall i\in\mathcal{N},\\
\nonumber
&\| C_i(k+1)-C_i(k)\| +
\left(r_i(k+1)+r_i(k)\right) \\
\label{eq:const_r_2}
&\qquad \qquad \le V_{\mathrm{max}},~\forall k\in\{t_i \cup \mathcal{T}_i\},~\forall i\in\mathcal{N},
\end{align}
where $r_i(t_i)=0$ and $r_i(T_i)=0$, $i\in \mathcal{N}$. 
We illustrate the meaning of \req{eq:const_r_1} and \req{eq:const_r_2} with \rfig{fig:r_const}.
Equation \req{eq:const_r_1} means that the minimum distance of every pair of adjacent regions is longer than or equal to the \emph{realizable} minimum distance, i.e., the trajectory is realized for some $u_i$, $\psi_i$, $v_i$, $\theta_i$, $x_i$, and $y_i$ satisfying \req{eq:dynamics}-\req{eq:conflict}.
In a similar manner, equation \req{eq:const_r_2} means the maximum distance of every pair of adjacent regions is shorter than or equal to the realizable maximum distance. 
\end{itemize}
Finally, a constraint for ATM system operation is given.
\begin{itemize}
\item Operation Constraint\\
We define a constraint that works at re-planning of trajectories.
Consider that a trajectory planning is performed at a time $k = t_i$.
Then, we suppose that pilots select $\hat{C}_i(k)$, $k\in \mathcal{T}_i(t_i)$.
It should be noted that \emph{all} the future trajectories $\hat{C}_i(t_i+1)$ to $\hat{C}_i(T_i-1)$ are selected by pilots $i\in \mathcal{N}$ based on their aims.
At the re-planning, we incorporate the aims into trajectories to be redesigned by ATC.
Let us consider that trajectories are re-planed at a time $t_i+\tau$, where $\tau \in \{1, 2,\ldots, T_i -2 -t_i \}$.
Then, we need to impose a constraint on trajectories redesigned by ATC.
We let $C_{\mathrm{pre},i}(k) = \hat{C}_i(k)$, $k\in \mathcal{T}_i(t_i)$.
Then, trajectory-sets $\mathcal{W}_i(k)$, $k\in \{t_i+\tau +1,\ldots,T_i-1\}$ to be redesigned must include the last selection $C_{\mathrm{pre},i}(t_i+\tau +1)$ to $C_{\mathrm{pre},i}(T_i-1)$ such that pilots can select the same \emph{best} trajectories.
This constraint is described by a condition on $C_i(k)$ and $r_i(k)$ to be redesigned as 
\begin{align}
\label{eq:operation}
\| C_{\mathrm{pre},i}(k) - C_{i}(k) \|
\le
r_i(k),~\forall k\in \mathcal{T}_i(t_i+\tau _i),~\forall i\in\mathcal{N}.
\end{align}
The meaning of the equation is illustrated in \rfig{fig:operation}.
\end{itemize}


\subsection{Optimization Problem}
\textbf{Problem~1}, which is stated in Section I\hspace{-.1em}I, is mathematically formulated in this subsection.
To this end, we let $X:=[\, X_1^{\top} X_2^{\top} \cdots X_N^{\top}\,]^{\top}$.
Then, the optimization problem addressed by ATC is summarized as follows.
\begin{problem}
For given $X_i(t_i)$, $X_i(T_i)$, $C_{\mathrm{pre},i}(k)$, $k\in \mathcal{T}_i(t_i)$, $i\in \mathcal{N}$,
\begin{align}
&\mini _{u,\psi,r,X}~J_{\mathrm{ATC}}(u,\psi,r)\\
\nonumber
&\mathrm{subject~to~Eqs.}~\reqs{eq:dynamics}-\reqs{eq:operation}.
\end{align}
\end{problem}
In the following, the optimizer to \textbf{Problem 3} is denoted by $(u^{\ast}$, $\psi^{\ast}$, $r^{\ast}$, $X^{\ast})$.
A trajectory-set $\mathcal{W}^{\ast}_i(k)$, $k\in \mathcal{T}_i(t_i)$ is obtained based on the solution to \textbf{Problem~3}.
Then, $\mathcal{W}^{\ast}_i(k)$, $k\in \mathcal{T}_i(t_i)$ is provided to each pilot.

\begin{note}
Note again that \textbf{Problem~3} is formulated without utilizing information on detailed aircraft dynamics and weather conditions.
This is because that such information is not necessary for system management including the conflict avoidance.
On the other hand, the information can be efficiently utilized by pilots, e.g., for reducing fuel costs.
The optimization problem addressed by pilots utilizes aircraft models based on BADA data\cite{nuic2010bada} or weather situation based on global forecasting systems, e.g., GDPFS (The Global Data-Processing and Forecasting System)\cite{GDPFS}.
\end{note}

\begin{figure}
\centering
\includegraphics[scale=0.45]{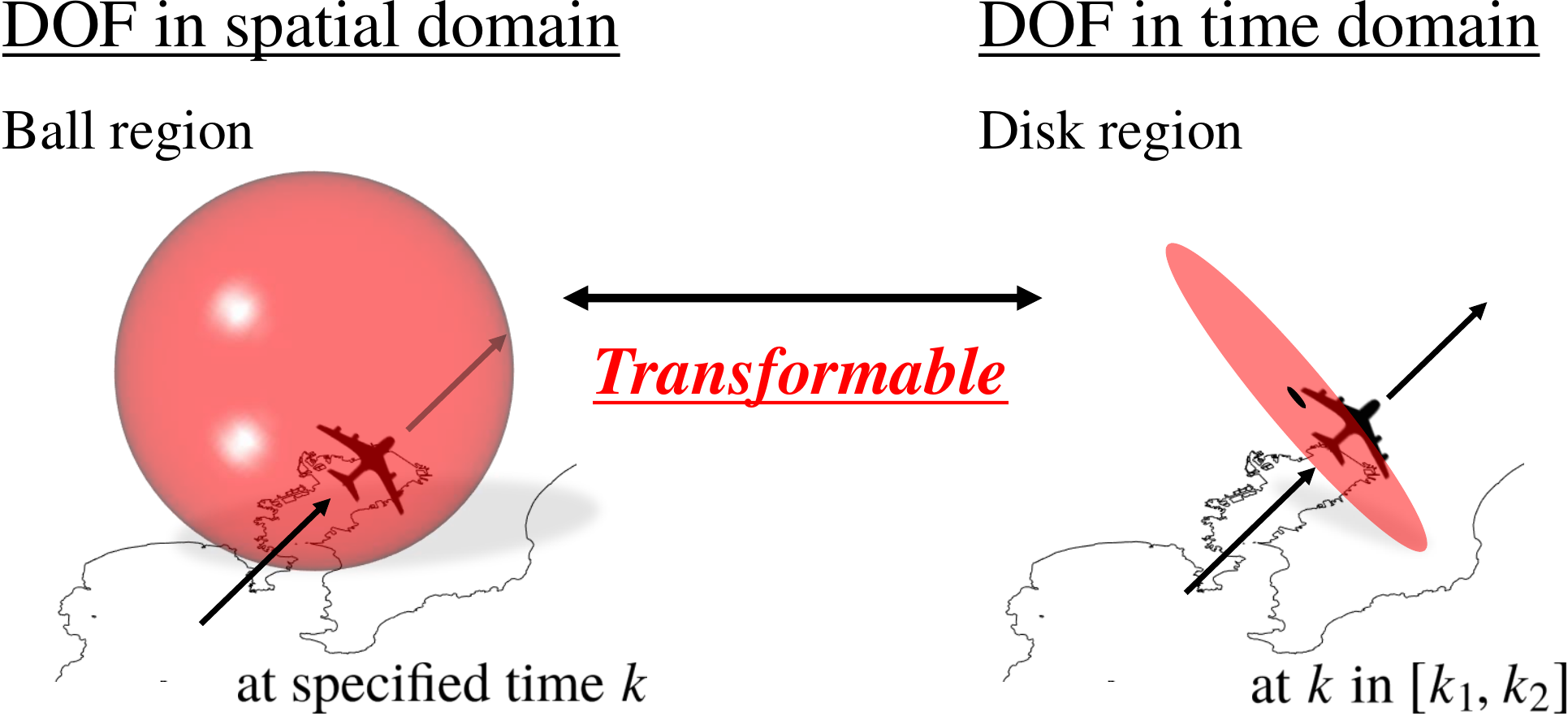}
\caption{Spatial DOF and temporal DOF.
A spatial DOF is a spatial ball region where an aircraft must be included at a \emph{ specified time}. 
A temporal DOF is a time range at \emph{any of which} an aircraft must pass a vertically-placed disk.
\label{illust_DOF}}
\end{figure}

\begin{note}
In this paper, a DOF in trajectories is expressed by spatial regions, denoted by $\mathcal{W}_i(k) = \{C_i(k),\,r_i(k)\}$.
It should be emphasized that the spatial DOF is equivalently transformed into temporal one.
As studied above, a spatial DOF in trajectories is expressed by a spatial ball region that aircraft must be included at a \emph{specified time} $k$.
On the other hand, a temporal DOF is expressed by a time range at \emph{any of which} aircraft passes a specified vertically-placed disk.
The expressions are illustrated in \rfig{illust_DOF}. 
The two expressions of DOF are equivalent each other.
\end{note}

\begin{note}
Since \textbf{Problem~3} is in the class of high-dimensional nonlinear optimization problems, the choice of initial values of decision variables is essential to obtain a \emph{better} local optimum.
In this paper, the initial guess of $u$ and $\psi$ is determined based on standard trajectories.
Otherwise, at re-planning case, the initial guess of $C_i(k)$, $k\in \mathcal{T}_i(t_i)$ is based on pilots selections $C_{\mathrm{pre}}(k)$, $k\in \mathcal{T}_i(t_i)$, which are designed at the last planning.
\end{note}

\section{Trajectory Design by Pilots}
In Section I\hspace{-.1em}I\hspace{-.1em}I, the trajectory design problem addressed by \emph{ATC} is formulated.
In Section I\hspace{-.1em}V, we focus on a trajectory design problem addressed by \emph{pilots}.
In the design problem of this section, we focus only on a pilot labeled by $i\in\mathcal{N}$.
It is assumed that the reference trajectory-set, denoted by $W_i^{\ast}(k)$, $k\in \mathcal{T}_i(t_i)$, is given and available for the design problem.
Then, in a similar manner to the ATC case, the design problem by the pilot is formulated as an optimization problem.
In this section, all decision variables of the optimization problem are denoted as $\hat{\cdot}$ in order to distinguish them from those in \textbf{Problem 3}. 

\subsection{Cost Function}
The cost function to evaluate the aims of the pilot is given.
The decision variables are speed difference $\hat{u}_i$ and angular difference $\hat{\psi}_i$.
The cost function is described by
\begin{align}
\label{eq:cost-pilot}
J_{\mathrm{pilot}}(\hat{u}_i,\hat{\psi}_i)
:=
\sum_{k=t_i}^{T_i-2}
\left(
\left( \dfrac{\hat{u}_i(k)}{U}\right)^2
+\left( \dfrac{\hat{\psi}_i(k)}{\Psi}\right)^2
\right).
\end{align}
This cost function $J_{\mathrm{pilot}}$ evaluates fuel costs of aircraft labeled by $i\in \mathcal{N}$.
The fuel costs are defined as a sum of squared control inputs of aircraft based on the definition of the work\cite{toratani2018effects}.
In \req{eq:cost-pilot}, variables $\hat{u}_i$ and $\hat{\psi}_i$ are normalized such that the effects of them are fairly evaluated. 

\subsection{Constraints}
Constraints that come from physical properties of aircraft and from ATM system requirements are listed.
Constraints from aircraft specification are the same as \reqs{const_angle}-\req{eq:velocity}, while those from aircraft dynamics are more precise than \req{eq:dynamics}.
We assume that information of disturbance $d$, which represents, e.g., wind condition of airspace, is available for trajectory design by the pilot.
Then, aircraft dynamics are modified from \req{eq:dynamics} to 
\begin{align}
\label{eq:dynamics-p}
\begin{bmatrix}
\hat{x}_i(k+1)\\
\hat{y}_i(k+1)\\
\hat{v}_i(k+1)\\
\hat{\theta}_i(k+1)\\
\end{bmatrix}
=
f\left(
\begin{bmatrix}
\hat{x}_i(k)\\
\hat{y}_i(k)\\
\hat{v}_i(k)\\
\hat{\theta}_i(k)\\
\end{bmatrix}
\right)
+
\begin{bmatrix}
0\\
0\\
\hat{u}_i(k)\\
\hat{\psi}_i(k)\\
\end{bmatrix}
+
\begin{bmatrix}
d_{x,i}(k)\\
d_{y,i}(k)\\
0\\
0\\
\end{bmatrix}.
\end{align}
In this problem setting, disturbances $d_i(k)$ are only applied to aircraft positions, in other words, $d_i(k)=[\,d_{x,i}(k)~d_{y,i}(k)~0~0\,]^{\top}$ holds.

For constraints from ATM system requirements, in addition to \reqs{eq:term_velocity} and \req{eq:term_angle} each pilot selects his/her own trajectory from the trajectory set, which is designed and requested by ATC.
Recall here that an aircraft trajectory to be designed by the pilot is denoted by $\hat{C}_i(k)$ and that the trajectory-set designed by ATC is denoted by $\mathcal{W}^{\ast}_i(k)=\{ C^{\ast}_{i}(k),r^{\ast}_i(k)\}$.
Then, the constraint on the trajectory selection is described by 
\begin{align}
\label{const_pilot}
r^{\ast}_i(k) &\ge \| C^{\ast}_{i}(k)-\hat{C}_i(k)\|_2,~\forall k\in \mathcal{T}_i(t_i).
\end{align}

\subsection{Optimization Problem}
The optimization problem addressed by each pilot is summarized as follows.
\begin{problem}
For given $X^{\ast}_i(t_i)$, $X^{\ast}_i(T_i)$, $\mathcal{W}^{\ast}_i(k)$, $k\in \mathcal{T}_i(t_i)$,
\begin{align}
\label{eq:opt-p}
&\mini_{\hat{u}_i,\hat{\psi}_i,\hat{X}_i}~J_{\mathrm{pilot}}(\hat{u}_i,\hat{\psi}_i)\\
\nonumber
&\mathrm{subject~to~Eqs.}~\reqs{eq:f}-\req{eq:term_angle},\reqs{eq:dynamics-p},\reqs{const_pilot}.
\end{align}
\end{problem}
In the following, the optimizer to \textbf{Problem 4} is denoted by $(\hat{u}^{\dagger}_i$, $\hat{\psi}^{\dagger}_i$, $\hat{X}^{\dagger}_i)$.
Then, the optimal trajectory is denoted by $\hat{C}^\dagger_i(k)$, $k\in \mathcal{T}_i(t_i)$.

\begin{figure}
\centering
\includegraphics[scale=0.4]{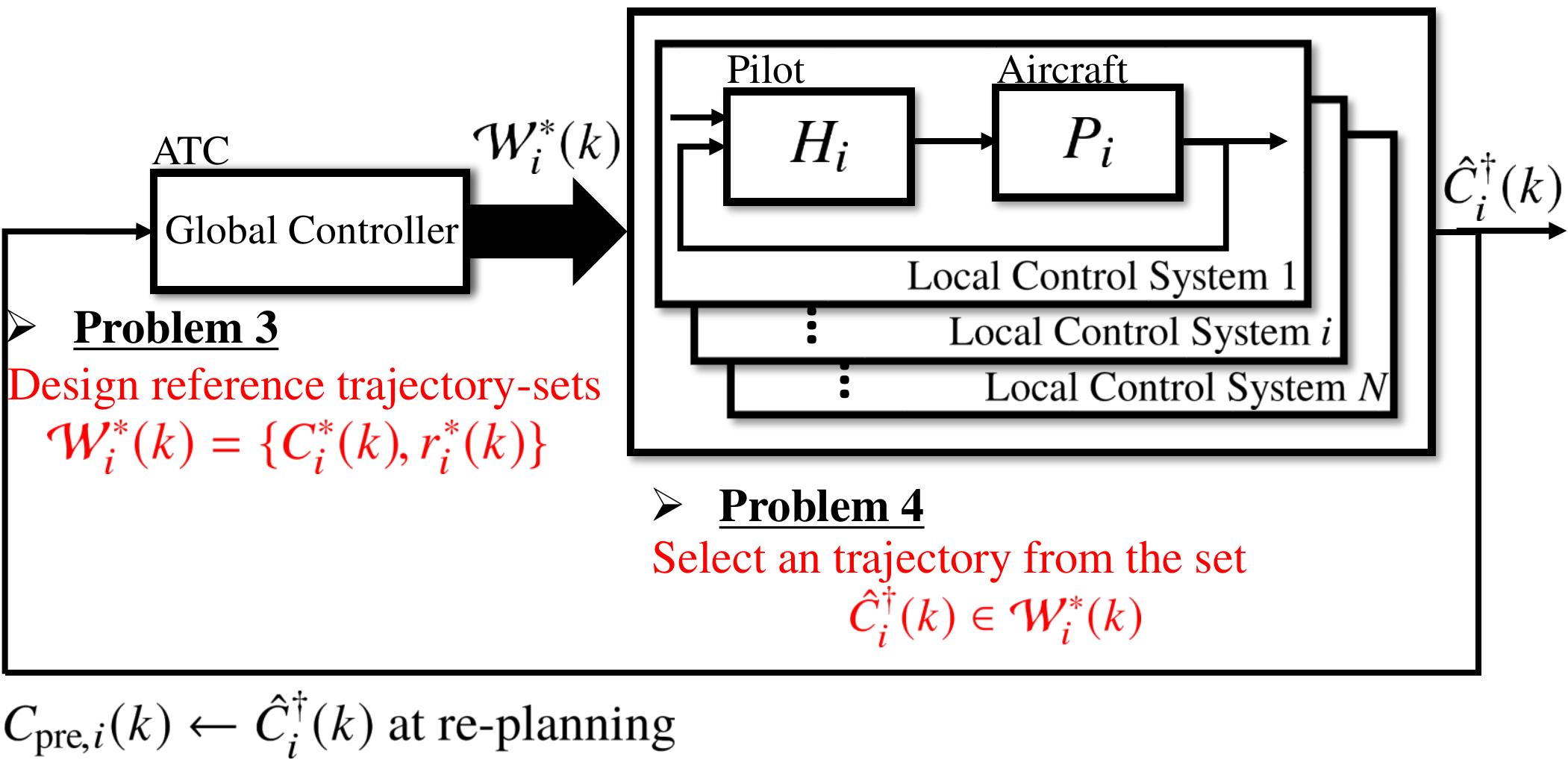}
\caption{Operation flow of the proposed ATM system.
Trajectory planning runs recursively at some time interval. 
In one trajectory planning, trajectory design is performed sequentially by ATC and pilots.
Trajectory-sets and trajectories are determined based on the solutions to \textbf{Problem~3} and \textbf{4}, respectively.
\label{concept_detailed}}
\end{figure}

\begin{note}
In this paper, no precise model of aircraft dynamics is utilized even in \textbf{Problem~4}.
In a practical setting, \req{eq:dynamics-p} need to be replaced by a more precise model.
Utilizing more detailed aircraft dynamics, the problem is formulated in the same manner.
\end{note}

The whole operation flow of the proposed ATM system is reviewed.
The block diagram shown as \rfig{concept_detailed} illustrates a more detailed operation flow than that of \rfig{concept}.
Each trajectory planning is composed of two stages; 1) trajectory-set design by ATC and 2) trajectory design by pilots.
1) ATC designs trajectory-sets by solving \textbf{Problem~3}.  
In the problem setting of \textbf{Problem~3}, given initial states $X_i(t_i)$, terminal states $X_i(T_i)$, trajectories selected by pilots at the last trajectory planning, denoted by $C_{\mathrm{pre},i}(k)$, $k\in \mathcal{T}_i(t_i)$, $i\in \mathcal{N}$, we aim at finding reference trajectory-sets $\mathcal{W}_i(k)$, $k\in \mathcal{T}_i(t_i)$.
ATC requests pilots to select their trajectories from the designed sets $\mathcal{W}_i^{\ast}(k)$, $k\in \mathcal{T}_i(t_i)$.
2) Then, each pilot design his/her trajectory by solving \textbf{Problem~4} independently of each other.
In the problem setting of \textbf{Problem~4}, given $X_i^{\ast}(t_i)$, $X_i^{\ast}(T_i),~i\in \mathcal{N}$, we aim at finding trajectories $\hat{C}_i(k)$, $k\in \mathcal{T}_i(t_i)$.
If re-planning is required due to some real-world uncertainties, we let $C_{\mathrm{pre},i}(k)=\hat{C}^{\dagger}_i(k)$, $k\in \mathcal{T}_i(t_i)$.
Then, the two stage design is performed again. 

\section{Numerical Simulations}
\begin{figure}
\centering
\includegraphics[scale=0.6]{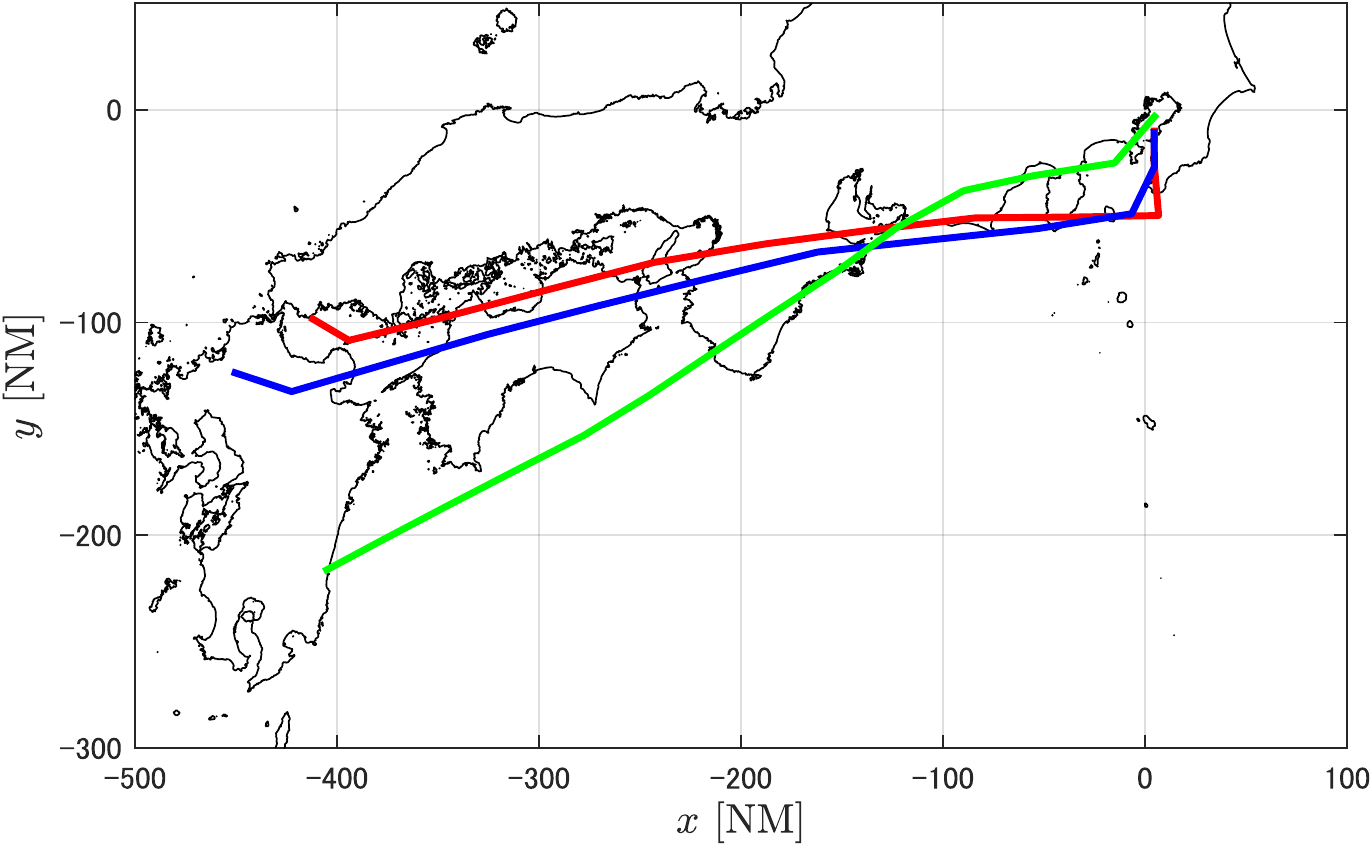}
\caption{Actual trajectories are extracted from CARATS Open Data.
The aircraft trajectory that departs from HANEDA is represented by red and blue solid lines, while that arrives at HANEDA is represented by the green solid line.\label{fig:actual_path}}
\end{figure}

We verify the effectiveness of the proposed ATM system, including aircraft trajectory design, in a numerical simulation.
To this end, the proposed design method is applied to actual data extracted from CARATS Open Data.
The data includes three trajectories from or arrive at HANEDA airport on May 11, 2015.
These trajectories are illustrated in \rfig{fig:actual_path}.
In the figure, HANEDA airport is located at the origin.
The aircraft trajectories that depart from HANEDA are represented by red and blue solid lines, while that arrives at HANEDA is represented by the green solid line.

\subsection{Simulation Conditions}
In the simulation, trajectory planning runs at the initial time and middle time of the operation.
At the middle time, only trajectory design by ATC is demonstrated in this section.
Conditions on the first and second trajectory planning are given below.

Conditions on the first trajectory planning are given.
For each aircraft $i\in \mathcal{N}=\{1,2,3\}$, the initial operation time $t_i$, the terminal operation time $T_i$, the initial state $X_i(t_i)$, and the terminal state $X_i(T_i)$ are given as follows.
\begin{itemize}
\item Aircraft 1
\begin{align*}
&t_1=1,~T_1=12,~X_1(t_1)=[\,4.71~-8.42~16.4~-1.58\,]^{\top}\\
&X_1(T_1)=[\,-413~-97.5~22.2~2.63\,]^{\top}
\end{align*}
\item Aircraft 2
\begin{align*}
&t_2=2,~T_2=13,~X_2(t_2)=[\,4.50~-9.01~17.7~-1.56\,]^{\top}\\
&X_2(T_2)=[\,-452~-123~31.1~2.84\,]^{\top}
\end{align*}
\item Aircraft 3
\begin{align*}
&t_3=2,~T_3=15,~X_3(t_3)=[\,-406~-217~31.8~0.471\,]^{\top}\\
&X_3(T_3)=[\,5.91~-1.96~31.2~0.833\,]^{\top}
\end{align*}
\end{itemize}
Here, one time step corresponds to six minutes.

In the cost function \req{eq:atc_cost}, the parameter is chosen by $\alpha = 0.01$.
To solve \textbf{Problem~3}, the initial values of $u$ and $\psi$ are computed based on the standard trajectories in actual data, while that of $r$ is zero. 

In \textbf{Problem~4}, for simplicity, we suppose that the wind disturbance $d$ is constant and that the magnitude and angle of $d$ are $1/3$ and $\pi/4$, respectively. 
That is, $d_{x,i}(k)=0.236,~d_{y,i}(k)=0.236,~k\in \mathcal{T}_i,~i\in \mathcal{N}$.
Note that only pilots utilize this wind condition.
To solve \textbf{Problem~4}, the initial values of $\hat{u}$ and $\hat{\psi}$ are determined such that the center of the trajectory-sets requested by ATC are tracked. 

Conditions on the second trajectory planning are given.
The second trajectory planning is carried out at the fifth time step, i.e., $t_i=5,~i\in\mathcal{N}$.
The parameter $\alpha$ of \req{eq:atc_cost} is fixed as $\alpha = 0.01$, which is the same as the first planning.
To solve the problem, the initial values of $\hat{u}$ and $\hat{\psi}$ are the same as ones selected by pilots at the first trajectory planning. 
In addition, the initial value of $\hat{r}$ is zero.

\subsection{Results}
\begin{figure}[H]
\centering
\includegraphics[scale=0.6]{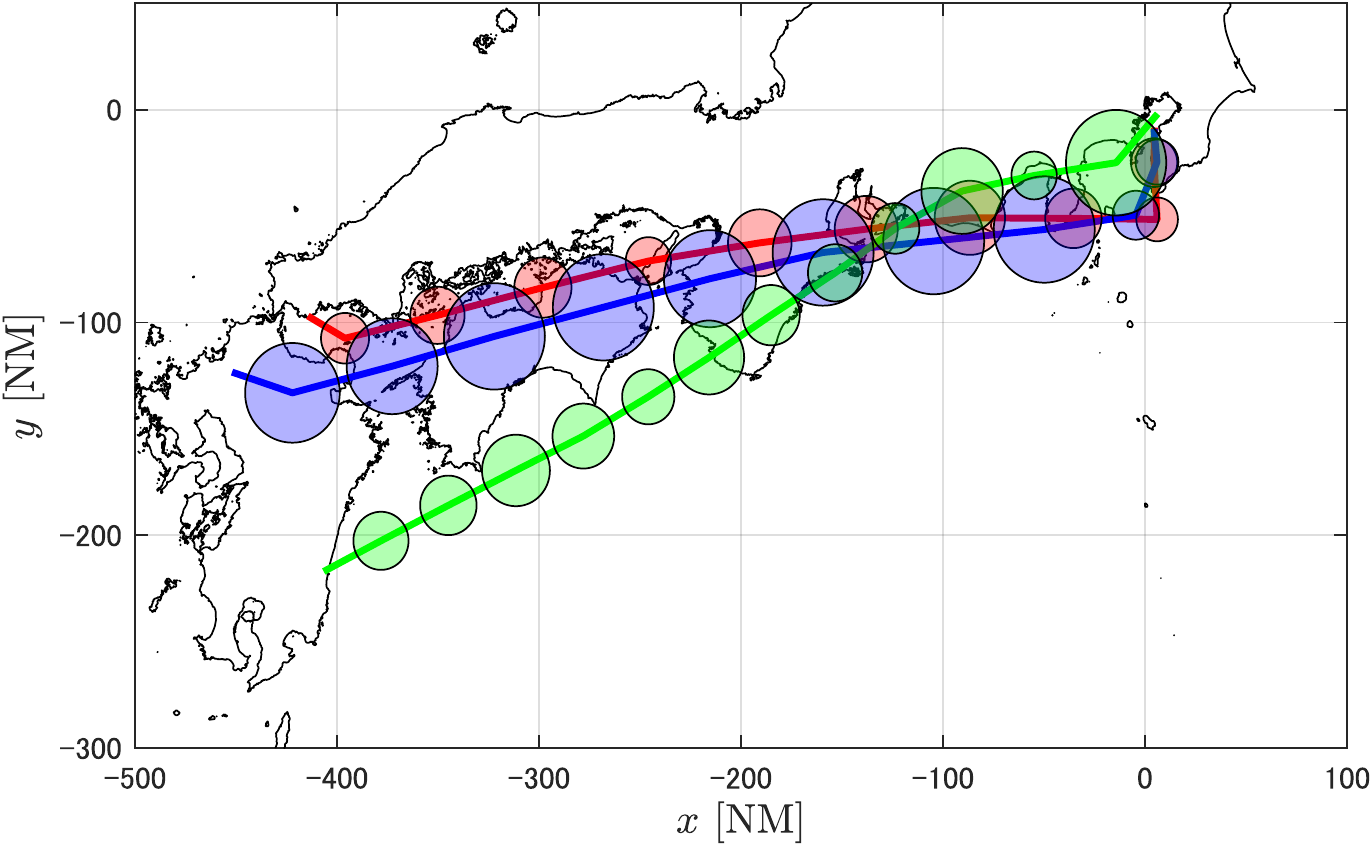}
\caption{Result of ATC design at the first trajectory planning.
The sets designed by ATC are represented by red, blue, and green disks.
Center of the disks is connected and represented by the solid lines.
\label{fig:atc_traj}}
\vspace{3mm}
\includegraphics[scale=0.6]{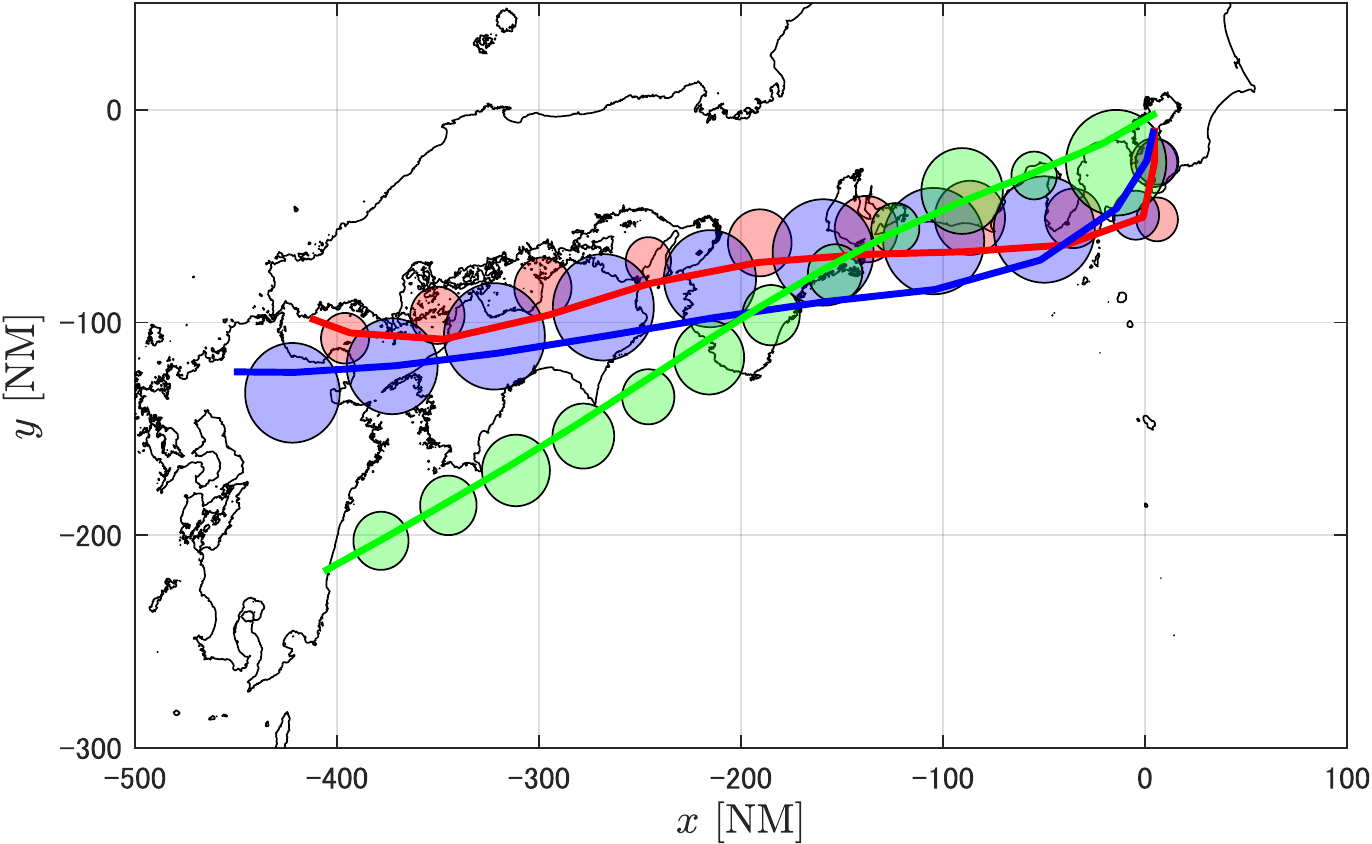}
\caption{Result of pilots design at the first trajectory planning.
The trajectories selected by pilots are represented by the solid lines.
The red, blue, and green disks are the sets designed and requested by ATC, which are the same as those of \rfig{fig:atc_traj}. 
\label{fig:pilot_traj}}
\vspace{3mm}
\includegraphics[scale=0.6]{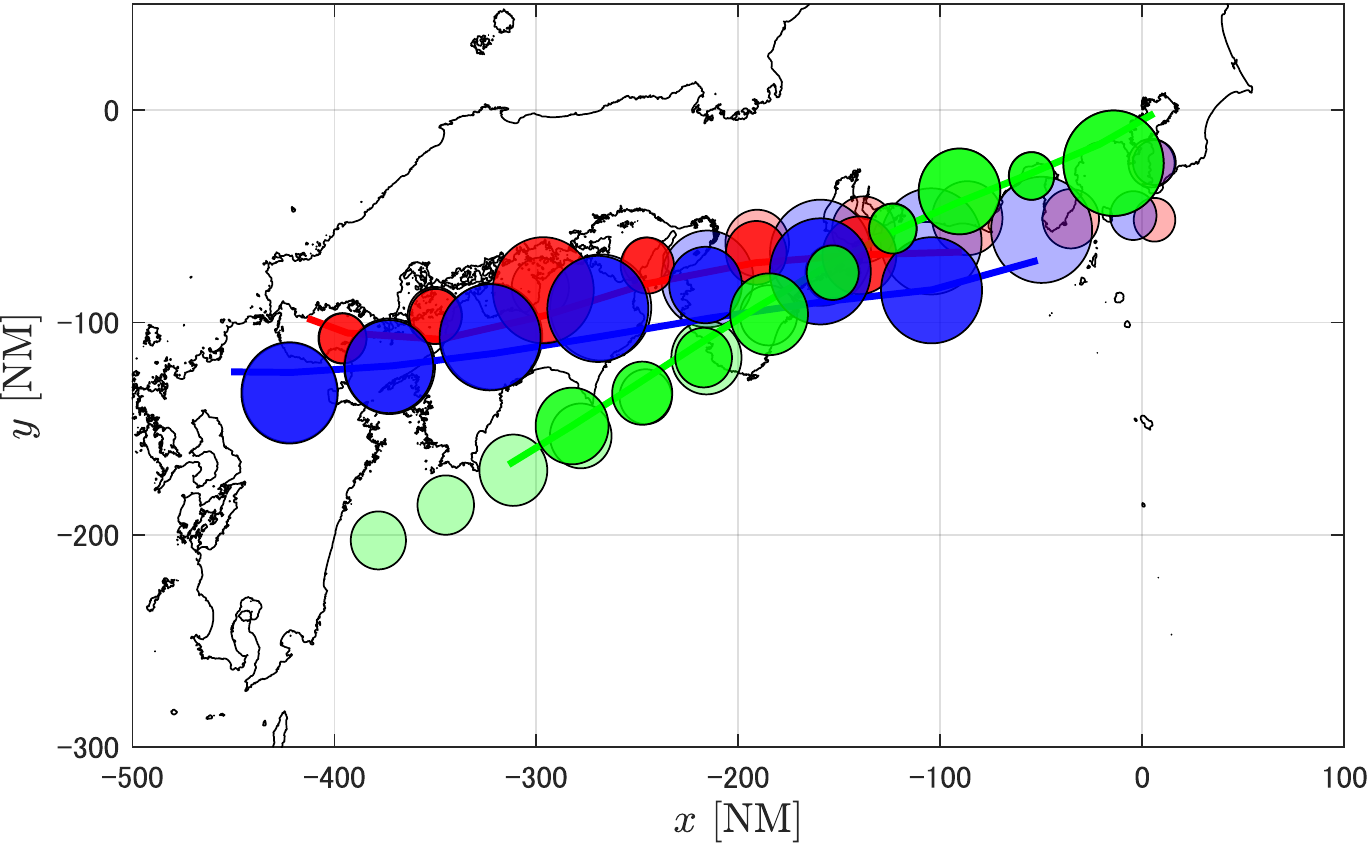}
\caption{Result of ATC re-design at the second trajectory planning. 
The sets redesigned by ATC are represented by deep red, blue, and green disks.
The trajectories selected by pilots at the first trajectory planning are represented by the solid lines.
\label{fig:atc_ope_traj}}
\end{figure}

The resulting trajectories of the first trajectory planning are shown in \rfigs{fig:atc_traj} and \ref{fig:pilot_traj}.
Figure \ref{fig:atc_traj} shows the result of the trajectory design by ATC.
In the figure, the reference trajectory-sets are represented by the red, blue, and green disks, while the trajectories that track the center of the sets are represented by the solid lines. 
The mean and standard deviation of the resulting $r_i(k)$, which is the DOF, are listed on \rtab{tab:r_res1}.
The DOFs of Aircraft 1 and Aircraft 2, which are adjoining, are the minimum and maximum one.
This is because that the cost tends to be small when one DOF in adjoining trajectories is big and the other is small.

Figure \ref{fig:pilot_traj} shows the result of the trajectory designed by pilots.
In the figure, the trajectories selected by pilots are represented by the red, blue, and green solid lines.
We see that all the trajectories are included in the trajectory-sets requested by ATC, which are represented by the disks.
This implies that conflict avoidance is achieved.
The trajectories are quantitatively evaluated by the cost function $J_\mathrm{pilot}$, which is defined in \req{eq:cost-pilot} and simulates fuel costs in some sense.
The cost values of actual trajectories (Trajectory A), trajectories that track the center of the sets (Trajectory B), and the trajectories selected by pilots (Trajectory C) are listed on \rtab{tab:cost_value}.
This shows that the fuel costs of Trajectory C is the minimum for all aircraft.
This result implies that although only a simple aircraft model and a performance index of fuel costs are used in this simulation, it is expected that performances are improved even when more detailed models and performance indexes are used.

\begin{table}
\centering
\caption{Comparison at the resulting DOF, defined by $\|r(k)\|$, at the first trajectory planning.\label{tab:r_res1}}
\begin{tabular}{lccc}\toprule
&Aircraft 1&Aircraft 2&Aircraft 3 \\ \hline \hline
Mean&$13.5$&$21.6$&$15.5$ \\ \hline
Standard deviation&$2.30$&$5.65$&$3.85$ \\ \bottomrule
\end{tabular}
\vspace{3mm}
\caption{Comparison at the values of \req{eq:cost-pilot} as fuel costs.\label{tab:cost_value}}
\begin{tabular}{lcccc}\toprule
&Aircraft 1&Aircraft 2&Aircraft 3&total \\ \hline \hline
Trajectory A&$2.91$&$1.95$&$0.44$&$5.31$ \\ \hline
Trajectory B&$2.99$&$2.07$&$0.47$&$5.44$ \\ \hline
Trajectory C&$2.13$&$1.62$&$0.25$&$4.00$ \\ \bottomrule
\end{tabular}
\end{table}

Figure \ref{fig:atc_ope_traj} shows the result of the second trajectory planning.
In the figure, the reference trajectory-sets redesigned by ATC are drawn on \rfig{fig:pilot_traj}.
The sets are represented by the deep red, blue, and green disks.
It is confirmed that the constraint for operation is satisfied; the trajectories selected by pilots at the first trajectory planning are included in the trajectory-sets redesigned by ATC.
This implies that each pilot can choose the same trajectory as the last planning if there is no update for the weather condition and pilots and airlines cannot receive demerits by the re-planning.

\clearpage

\section{Conclusion}
In this paper, we proposed a novel framework of ATM systems where ATC \emph{weakly supervises} aircraft.
Aircraft trajectories, which are completely determined by ATC conventionally, are
designed by ATC as \emph{trajectory-sets}, and the sets are provided to pilots.
Then, the pilots individually select their trajectories from the sets.
We showed that both the safety requirement, which is the aim of ATC, and reduction of fuel consumption, which is the aim of pilots and airlines, were achieved in the ATM system.
The authors expect that the proposed ATM system may not be directly implemented to real-world ATM.
However, the essence of the weak supervision, i.e., idea of explicitly providing DOFs to pilots, can be utilized to practical ATM systems in a modified manner.

The optimization problems formulated for trajectory planning are nonlinear to some decision variables, which are numerically intractable.
In future work, we need to reformulate them to more tractable ones.

\section*{Acknowledgment}
This work was supported by Grant for Basic Science Research Projects from Sumitomo Foundation and by Grant-in-Aid for Scientific Research (A), No.~18H03774 from JSPS.

\ifCLASSOPTIONcaptionsoff
  \newpage
\fi



%
%
%

\bibliographystyle{IEEEtran}
\bibliography{reference}

%

\begin{IEEEbiography}{Sho Yoshimura}
Sho Yoshimura was born in Chiba, Japan in 1993.
He received the Bachelor's degree in Engineering from Keio University in 2017.
He is currently a Master Course student in Keio University.
His research interests include air traffic management design and system identification for large-scale systems.
\end{IEEEbiography}

\begin{IEEEbiography}{Masaki Inoue}
Masaki Inoue was born in Aichi, Japan, in 1986.
He received the M.E. and Ph.D. degrees in engineering from Osaka University in 2009 and 2012, respectively.
He served as a Research Fellow of the Japan Society for the Promotion of Science from 2010 to 2012.
From 2012 to 2014, He was a Project Researcher of FIRST, Aihara Innovative Mathematical Modelling Project, and also a Doctoral Researcher of the Graduate School of Information Science and Engineering, Tokyo Institute of Technology.  Currently, he is an Assistant Professor of the Faculty of Science and Technology, Keio University.
His research interests include stability theory of dynamical systems.
He received several research awards including the Best Paper Awards from SICE in 2013, 2015, and 2018, from ISCIE in 2014, from IEEJ in 2017, and the Takeda Best Paper Award from SICE in 2018.
He is a member of IEEE, SICE, ISCIE, and IEEJ.
\end{IEEEbiography}





\end{document}